\documentclass[12pt,a4paper]{article}
\usepackage{natbib,graphicx,amssymb,bm,setspace,dcolumn,amsmath,todonotes}
\usepackage{bbm,tabularx,url,multirow}

\renewcommand\footnotemark{}

\title{An efficient semiparametric maxima estimator of the extremal index
}
\author{Paul J. Northrop \\
\footnote{
{\it Address for correspondence}: Paul Northrop, Department of Statistical Science, University College London, Gower Street, London WC1E 6BT, UK. 
E-mail: p.northrop@ucl.ac.uk}
{\it University College London, UK}  
}

\newcommand{\disp}{\displaystyle}
\let\leq=\leqslant
\let\geq=\geqslant
\let\hat=\widehat
\newcommand{\bit}{\begin{itemize}}
\newcommand{\eit}{\end{itemize}}
\newcommand{\beqnn}{\begin{eqnarray*}}
\newcommand{\eeqnn}{\end{eqnarray*}}
\newcommand{\beqn}{\begin{eqnarray}}
\newcommand{\eeqn}{\end{eqnarray}}

\newcommand{\E}{{\rm E}}
\newcommand{\var}{{\rm var}}
\newcommand{\cov}{{\rm cov}}

\def\dperp{
\mathrel{%
\kern0pt\vbox{\hbox to 0.8em{\hss
\setbox0=\hbox{$\shortparallel$}\dp0=0pt\box0\hss}\hrule width 0.8em}%
}}

\begin{document}
\maketitle

\vspace{-1.2cm}

\begin{abstract}
The extremal index $\theta$, a measure of the degree of local dependence in the
extremes of a stationary process, plays an important role in extreme value analyses.
We estimate $\theta$ semiparametrically, using the relationship between the 
distribution of block maxima and the marginal distribution of a process to define a semiparametric model.  We show that these semiparametric estimators are simpler and substantially more efficient than their parametric counterparts.
We seek to improve efficiency further using maxima over sliding blocks.
A simulation study shows that the semiparametric estimators are competitive with the leading estimators.
An application to sea-surge heights combines inferences about $\theta$ with a standard extreme value analysis of block maxima to estimate marginal quantiles.
\end{abstract}


\vspace{-0.2cm}

\section{Introduction}
\label{sec:intro}
The modelling of rare events in stationary processes is important in many application areas. 
The extremal behaviour of such a process is governed by its marginal distribution and by its extremal dependence structure.
\cite{CD2012} provide a review of this area, concentrating on the latter aspect.
For processes satisfying the $D(u_n)$ condition of \cite{LLR1983}, which limits long-range dependence at extreme levels, the extreme value index $\theta \in [0,1]$ is the primary measure of short-range extremal dependence.

Following \cite{LLR1983}, let $X_1,X_2,\ldots$ be a strictly stationary sequence of random variables that satisfies the $D(u_n)$ condition and has marginal distribution function $F$.  Let $M_b=\max(X_1, \ldots, X_b)$.   
In the non-degenerate case when $\theta>0$,
for large $b$ and $u_b$ the distribution function $G_b$ of the block maximum $M_b$ is approximately related to $F$ via
\begin{equation}
G_b(u_b)=P(M_b \leq u_b) \approx F^{b\theta}(u_b). 
\label{eqn:main_equation}
\end{equation}
Further, if there exist normalizing constants $c_b$ and $d_b$ such that
$F^b(c_bx+d_b) \rightarrow G(x)$, as $b \rightarrow \infty$,
then $G(x)$ is the distribution function of a Generalized Extreme Value (GEV) distribution.
The corresponding result for $M_b^*=\max(X_1^*,\ldots,X_n^*)$, where $X^*_1,X^*_2,\ldots$ are independent variables with distribution function $F$, gives the limiting distribution function 
$H(x)=G(x)^{1/\theta}$. 
Thus, the limiting distributions of $M_b$ and $M_b^*$ are GEV, with respective location, scale and shape parameters $(\mu_\theta,\sigma_\theta,\xi)$ and $(\mu,\sigma,\xi)$, say, related by 
\begin{equation}
\mu_\theta = \mu+\sigma \left( \theta^\xi-1\right)/\xi, \quad
\sigma_\theta =  \sigma \theta^\xi.
\label{eqn:rel}
\end{equation}
We consider a block size dependent index \citep{Smith1992} $\theta_b = -\log G(u_b)/\log 2$, where, without loss of generality, $F^b(u_b)=1/2$.
Thus, $\theta_b \rightarrow \theta$ as $b \rightarrow \infty$.
We will only make explicit the dependence of $\theta$ on $b$ when necessary.
\cite{ANT2000} (threshold dependent index $\theta(u)$) and \cite{RSF2009} (block size and threshold dependent index $\theta_b(u_b)$) consider similar sub-asymptotic forms.  Unfortunately no general theory exists concerning the rate of convergence to $\theta$ of these quantities.
Some of the bias in estimating $\theta$ is because $\theta_b$ (or $\theta(u)$ or $\theta_b(u_b)$), rather than $\theta$, is estimated.  

As noted by \cite{BGST2004}, ignoring $\theta$ leads to (a) underestimation of marginal quantiles of $F$ implied by inferences about $G$ from block maxima, and (b) overestimation of quantiles of $G$ implied by inferences about $F$ from, for example, a threshold-based analysis of raw data.
\cite{CD2012} note that $\theta$ contains information about 
the extent of clustering of extreme events that  may be of great practical importance.

\begin{sloppypar}
Recent advances in the estimation of $\theta$ \citep{FS2003,Suveges2007,SD2010,LT2003,Robert2013} have concentrated on {\it threshold methods}, based on exceedences of a threshold.
The improvement of {\it maxima methods} \citep{Gomes1993,ANT2000}, based on (\ref{eqn:rel}) and described in Section \ref{sec:maxima}, has received less attention.
We propose a new maxima method that is simpler and has much greater statistical efficiency than existing maxima methods.
\end{sloppypar}

\subsection{Maxima methods}
\label{sec:maxima}
 Parametric maxima methods are based on fitting GEV distributions to two sets of maxima of $b$ consecutive observations.  The first sample $M_i,i=1,\ldots,n$ is block maxima of the original series.  The second sample $M^*_i,i=1,\ldots,n$ is block maxima of a series obtained by randomizing the index of the original series, to obtain approximately an independent series with the same marginal distribution as the original sequence.  
Based on (\ref{eqn:rel}), \cite{Gomes1993} fit a GEV$(\mu_\theta,\sigma_\theta,\xi)$ distribution to $\{M\}$ and a GEV$(\mu,\sigma,\xi)$ distribution to $\{M^*\}$ and construct the estimator $\hat{\theta}_{G} = (\hat{\sigma}/\hat{\sigma}_\theta)^{-1/\tilde{\xi}}$, where $\tilde{\xi}=(\hat{\sigma}-\hat{\sigma}_\theta)/(\hat{\mu}-\hat{\mu}_\theta)$.
\cite{ANT2000} combine the two GEV fits into one by maximizing a likelihood (with respect to $(\mu,\sigma,\xi,\theta)$), assuming that ($\{M^*\}$, $\{M\}$) are independent.  We call the resulting estimator $\hat{\theta}_{AT}$.
In one sense parametric maxima methods are anomalous: other methods of estimating $\theta$ do so directly, without embedding $\theta$ in a larger model with nuisance parameters.
\cite{Northrop2005} proposes a {\it semiparametric} (SP) maxima estimator.  The relationship $G=H^\theta$ is used but no particular parametric form is assumed for $G$ or $H$.

An undesirable feature concerning existing maxima methods is the need to resample the original data to produce a sample of block maxima with approximate c.d.f. $H$.
In Section \ref{sec:SP} we show that this is unnecessary: more efficient estimators of $\theta$ can be constructed by comparing $G$ directly to $F$, without generating pseudo-samples from $H$.  
The theoretical gain in efficiency is quantified, albeit in an idealized situation, in Appendix A and in Section \ref{sec:desire} general properties of the semiparametric estimators are discussed.  
In Section \ref{sec:link} we show that one of these estimators is approximately an extended version of the {\it blocks estimator} of \cite{Robert2009}.
In Sections \ref{sec:sim} and \ref{sec:newlyn} we carry out simulation studies and an extreme value analysis of type 1 on sea-surge data respectively.
The paper is concluded in Section \ref{sec:discussion} with a discussion and technical proofs are reported in the Appendix.
Computer code to implement this methodology is available at \url{www.homepages.ucl.ac.uk/~ucakpjn/}.

\newcommand{\yd}{Y^{d}}
\newcommand{\ys}{Y^{s}}
\newcommand{\yid}{Y^{d}_i}
\newcommand{\yis}{Y^{s}_i}
\newcommand{\vid}{V^{d}_i}
\newcommand{\vis}{V^{s}_i}
\newcommand{\vidhat}{\hat{V}^{d}_i}
\newcommand{\vishat}{\hat{V}^{s}_i}

\section{Semiparametric maxima estimators of $\theta$}
\label{sec:SP}
Let $X_1,\ldots,X_m$ be strictly stationary sequence of random variables with marginal distribution function $F$ and extremal index $\theta$.  
Let $M(s,t]=\max_{s < k \leq t} X_k, n_d= \lfloor{m/b}\rfloor$ and $n_s=m-b+1$.
Consider two sets of block maxima: $\yd=\{\yid$, $i=1,\ldots,n_d\}$, where $\yid=M((i-1)b,ib]$ ({\it disjoint} blocks) and
$\ys=\{\yis,i=1,\ldots,n_s\}$, where $\yis=M(i-1,i+b-1]$ ({\it sliding} blocks).   
We use $n$ as general notation for the size of a sample of block maxima. 
Consider, for some $s \in \{0, \ldots, m-b\}$, $Y=M(s,s+b]$, the maximum of any block of $b$ consecutive $X$s,
and let $V=-b\log F(Y)$.
When $F$ is known and (\ref{eqn:main_equation}) holds then $V$ has an exponential distribution with mean $1/\theta$.
The maximum likelihood estimator (MLE) of $\theta$ based on a random sample $V_1,\ldots,V_n$ from this distribution is 
$\hat{\theta}_F=n/\sum_{i=1}^n V_i$, with variance ${\rm var}(\hat{\theta}_F)=n^2\theta^2(n-2)^{-1}(n-1)^{-2}$.

Typically $F$ is unknown, so we must use empirical analogues of $V$.
We describe these using the sliding block maxima $\ys$.
Let $V_i=-b \log F(Y_i^s), i=1, \ldots, m-b+1$ and 
let $B_i$ be the set of the $X$s that contribute to block maximum $Y_i^s$.
By construction, the $b$ values in $B_i$ cannot exceed block maximum $Y_i^s$.
To adjust for this deterministic effect we use only the $m-b$ values {\it not} in $B_i$ to construct the estimator $\hat{F}_{-i}$ of $F$ applied to $Y_i^s$.
Let $l_i=\min_{X_k \notin B_i}X_k$.
For $y \geq l_i$ we let
\begin{equation}
\hat{F}_{-i}(y) = \frac{1}{m-b+1} \sum_{X_k \notin B_i} \mathbbm{1}(X_k \leq y), 
\label{eqn:Fhat}
\end{equation}
where $\mathbbm{1}(A)$ is the indicator function of an event $A$.
$\hat{F}_{-i}(Y_i^s)$ can be expressed in terms of the rank $R_i \in \{1, \ldots, m-b+1\}$ of $Y_i^s$ within $X_1,\ldots,X_m$.
If $R_i=r_i$ then $r_i-1$ of $\{X_k, k \notin B_i\}$ are larger than $Y_i^s$ and $m-b+1-r_i$ of $\{X_k, k \notin B_i\}$ are smaller than $Y_i^s$.
Therefore, (\ref{eqn:Fhat}) gives $\hat{F}_{-i}(Y_i^s)=(m-b+1-R_i)/(m-b+1)$.
The case $R_i=m-b+1$ occurs only when $Y_i^s<l_i$, which is unlikely unless $b$ is small.  
To ensure positivity of $\hat{F}_{-i}(y)$ we set $\hat{F}_{-i}(y)$ to $1/(m-b+n+1)$ if $y < l_i$ \citep{DDM1989}.

Thus, $Y^s$ produces a sample $\hat{V}^s_i=-b \log \hat{F}_{-i}(Y_i^s), i=1, \ldots, n_s$, which are determined by the respective ranks $R_i, i=1, \ldots, n_s$.
The disjoint block maxima $Y^d$ produce the subsample $\hat{V}^d_i=\hat{V}^s_{(i-1)b+1}, i=1, \ldots, n_d$. 
We expect (as in \cite{RSF2009}) that $\ys$ contains more information about $\theta$ than $\yd$ and thus produces a more efficient estimator of $\theta$.

Consider block maxima $Y=(Y_1, \ldots, Y_n)$ with order statistics $Y'$ and
let $\hat{V}=\{\hat{V}_i, i=1,\ldots,n\}$, where $\hat{V}_i=-b\log \hat{F}_{-i}(Y_i)$.
The ranks $R=(R_1, \ldots, R_n)$ of $Y$ within $X_1, \ldots, X_m$ convey no information about the distribution of $Y'$.
Therefore, using a marginal GEV($\mu_\theta,\sigma_\theta,\xi$) model for the ordered block maxima $Y'$, the joint likelihood based on $Y=(\hat{V},Y')$ factorises as
\begin{equation}
L(\theta,\mu_\theta,\sigma_\theta,\xi;Y) = L_R(\theta;\hat{V}) \, L_{GEV}(\mu_\theta,\sigma_\theta,\xi;Y'), 
\label{eqn:joint}
\end{equation}
so that independent inferences can be made about $\theta$ and $(\mu_\theta,\sigma_\theta,\xi)$.
We use as an approximation to $L_R(\theta;\hat{V})$, the pseudo-likelihood 
\begin{equation}
L_{exp}(\theta;\hat{V})=\theta^n \exp\left(-\theta\sum_{i=1}^n \hat{V}_i\right),
\label{eqn:pseudo}
\end{equation}
that is, the likelihood that would apply if $\hat{V}_1, \ldots, \hat{V}_n$ are sampled randomly from an exponential distribution with mean $1/\theta$.   
The disjoint and sliding blocks estimators for $\theta$ are those that maximize the pseudo-likelihood $L_{exp}(\theta;\hat{V})$, that is,
\begin{equation}
\hat{\theta}_d = \left(\frac1{n_d}\sum_{i=1}^{n_d} \hat{V}^d_i \right)^{-1}, \quad 
    \hat{\theta}_s = \left(\frac1{n_s}\sum_{i=1}^{n_s} \hat{V}^s_i \right)^{-1}. 
\label{eqn:ests}
\end{equation}
For convenience we will use $\hat{\theta}_{SP}$ to refer to a general estimator of this type.

The pseudo-likelihood (\ref{eqn:pseudo}) is approximate because (\ref{eqn:main_equation}) provides only an approximate relationship between $G$ and $F$.
Serial dependence in the underlying sequence, $X_1, \ldots, X_m$ is expected, resulting in dependence between the values of $\hat{V}$ from nearby disjoint blocks.
Use of sliding blocks further complicates matters as successive values of $\hat{V}^s$ are strongly positively associated.
Even if (\ref{eqn:main_equation}) holds and the underlying sequence is i.i.d., estimation of $F$ from a finite sample introduces a further approximation.
Moreover, the double use of the sample to estimate $F$ and to provide block maxima, induces dependence between $\hat{V}_1, \ldots, \hat{V}_n$ that is not negligible asymptotically, see for example, \citet{Robert2009}. 
Therefore, the expression given for ${\rm var}(\hat{\theta}_F)$ at the start of Section \ref{sec:SP} does not apply in either case.

We attempt to make some adjustment for these issues by basing estimates of uncertainty on information sandwich estimators \citep{White1982} of the sampling variances of the estimators of $\theta$.   Details are given in Appendix B.
Some unrealistic simplifying assumptions are used, such as observations from distinct blocks being independent, so we use as an alternative a block bootstrap \citep{PR1994}.
In common with other estimators of $\theta$, studying the asymptotic properties of $\hat{\theta}_d$ and $\hat{\theta}_s$ is difficult
and we do not attempt such an analysis here.

\subsection{Desirable properties of these estimators}
\label{sec:desire}
The estimator $\hat{\theta}_{SP}$ is simple and non-iterative.  
Appendix A shows that, in an idealized situation where data can be treated as random samples from the respective models, it is more efficient than its parametric counterparts.  
This finding is supported by a simulation study presented in Section 3.
The extremal index measures local dependence in extremes and is independent of the marginal distribution of the process.  
Since $\hat{\theta}_{SP}$ is determined by the ordering of the ranks of the raw data it is invariant to marginal transformation, whereas the parametric alternatives are not.
In common with threshold methods, 
$\theta$ is estimated directly, rather than as part of a larger extreme value model,
that is, estimation of extremal dependence and marginal behaviour are separated.
It may be that the assumption ($G=F^{b\theta_b}$) underlying $\hat{\theta}_{SP}$ is reasonable for smaller block sizes than the parametric GEV assumptions.
Section \ref{sec:sim} gives examples where the rate of convergence of $\theta_b$ to $\theta$ as $b \rightarrow \infty$ is $O(1/b)$.
Thus, convergence to $G=F^{b\theta}$ could be relatively fast even if convergence to the limiting GEV form, which depends on the marginal distribution, is slow.
The semiparametric framework permits the use of a relatively small block size, whereas the parametric alternatives do not.

\subsection{Link to the blocks estimator}
\label{sec:link}
\begin{sloppypar}
The {\it blocks estimator} (sometimes called the {\it logs estimator}) $\hat{\theta}_B=\log\hat{G}(u) / b\log\hat{F}(u)$ \citep{SW1994} requires the choice of a threshold $u$ and a block size $b$. 
\cite{Robert2009} extends this idea by using a random threshold, set at a particular sample quantile.
We show that in large samples $\hat{\theta}_{d}$ gives approximately the same value as a combination of blocks estimators, each based on its own local data-dependent threshold.
Suppose that we use a set of random thresholds $u_i=Y_i$, where $Y_i, i=1,\ldots, n$ are a sample of block maxima over blocks of length $b$.  
We may think of this as using the data to define a set of local thresholds.
Each of the ratios $\log\hat{G}(Y_i) / b \log\hat{F}(Y_i)$ is an estimator of $\theta$.  
We combine these using a ratio estimator
\begin{equation}
\hat{\theta}_{RB} = \frac{\disp\frac1n\sum_{i=1}^n \log\hat{G}(Y_i)}{\disp b\, \frac1n\sum_{i=1}^n \log\hat{F}(Y_i)}
=\frac{\disp\frac1n\sum_{i=1}^n \log\hat{G}(Y_i)}{\disp -\frac1n\sum_{i=1}^n V_i}. \label{eqn:random_block}
\end{equation}
If disjoint blocks are used to construct $\{Y_i\}$ then $\hat{G}(Y_{(i)})=i/n$ and, by Stirling's formula, as $n \rightarrow \infty$ the numerator of (\ref{eqn:random_block}) $\downarrow -1$.
Therefore, for sufficiently large $n$, $\hat{\theta}_{d} \approx \hat{\theta}_{RB}$.
A potential advantage of maxima estimators over blocks estimators is that all block maxima contribute information directly to the estimator, regardless of whether or not they exceed some threshold $u$.
\end{sloppypar}

\subsection{Theoretical comparison with parametric maxima method}
\label{sec:comp}
\begin{figure}[h]
\centering
\includegraphics[width=0.75\textwidth, angle=0]{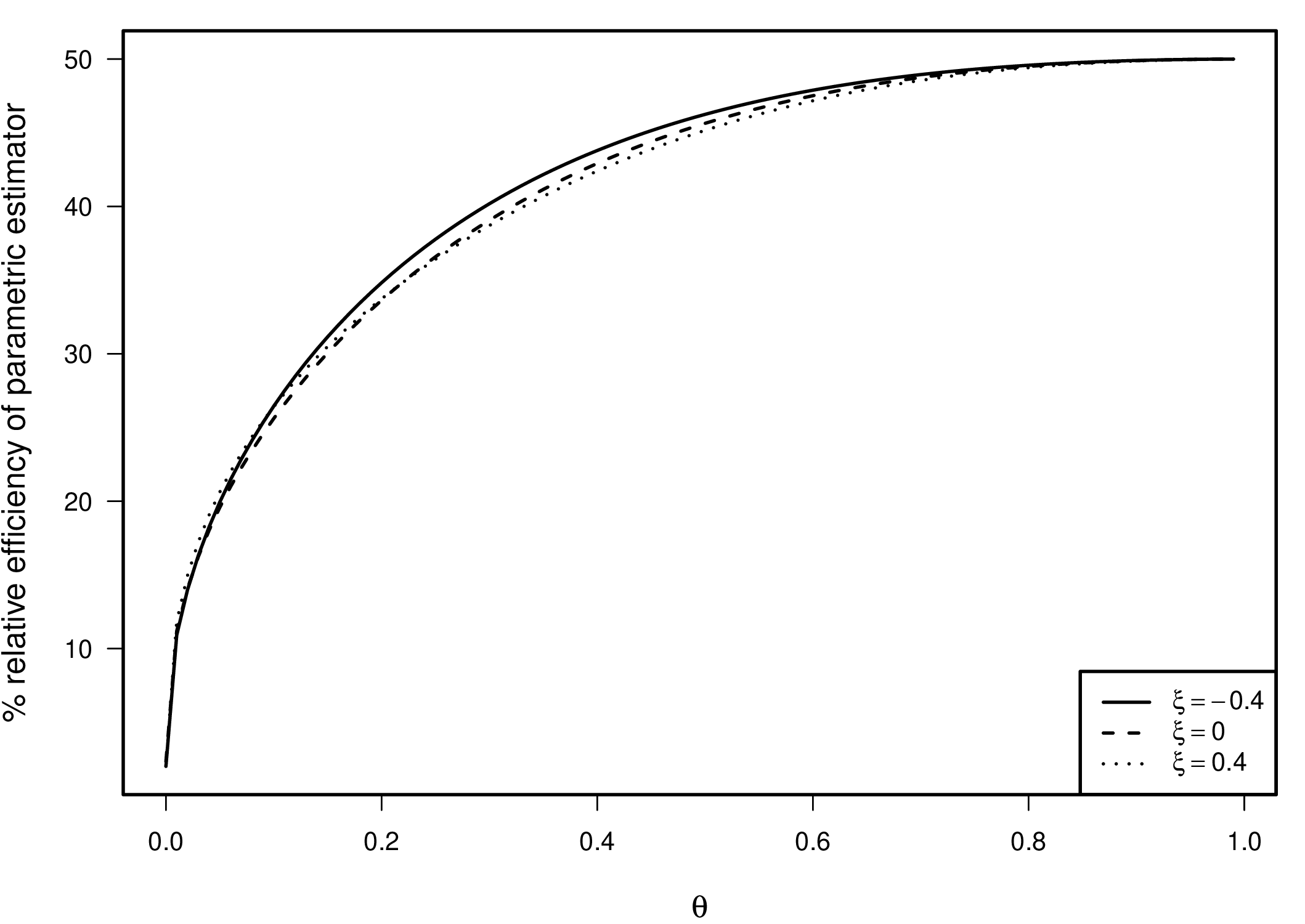}
\vspace*{-0.5cm}
\caption{\label{fig:rel_eff}
Asymptotic relative efficiency of the parametric maximum likelihood estimator of $\theta$ compared to the semiparametric maximum likelihood estimator for $\xi=-0.4$ (solid lines), $\xi=0$ (dashed line) and $\xi=0.4$ (dotted line).}
\end{figure}

The calculations in Appendix A show that, in an idealized situation where data can be treated as random samples from the respective models,  
$\hat{\theta}_{d}$ has a smaller asymptotic variance than $\hat{\theta}_{AT}$.
The asymptotic variance of $\hat{\theta}_{AT}$ depends on the marginal distribution of the raw data, via the shape parameter $\xi$ of the GEV distribution assumed for block maxima, whereas the asymptotic variance of $\hat{\theta}_{d}$ does not.
The efficiency of $\hat{\theta}_{AT}$ relative to $\hat{\theta}_{d}$ depends on $\theta$ and, to a lesser extent, on $\xi$ (see Figure \ref{fig:rel_eff}) 
but $\hat{\theta}_{AT}$ is at best 50\% efficient (when $\theta=1$).  
This is expected because the resampling used to produce the parametric estimator introduces an extra source of variability.

\section{Simulation studies}
\label{sec:sim}
We present two types of simulation study.   The first shows that the conjectured
superiority of the SP estimators relative to existing maxima methods is realised in practice and examines how best to estimate the sampling variability of the former.
The second compares the performance of the SP estimators to the most efficient threshold methods.

\subsection{Maxima estimators}
\label{sec:max_sim}
We compare the SP estimators of $\theta$ to the parametric estimators $\hat{\theta}_{AT}$ and $\hat{\theta}_G$ for different processes, values of $\theta$, marginal distributions and blocks sizes.  
The processes are those for which results are presented in Tables 4-8 of \cite{ANT2000}.
As the general findings are the same in all cases we present results only for a max-autoregressive (maxAR) process \citep{DR1989}: $X_i = \max \{ (1-\theta) X_{i-1}, \theta Z_i \}$, where $\{Z_i\}$ and $X_0$ have independent unit Fr\'{e}chet distributions.  
For this process $\theta_b=\theta+(1-\theta)/b$ (see Appendix C).

\begin{figure}[h]
\centering
\includegraphics[width=\textwidth, angle=0]{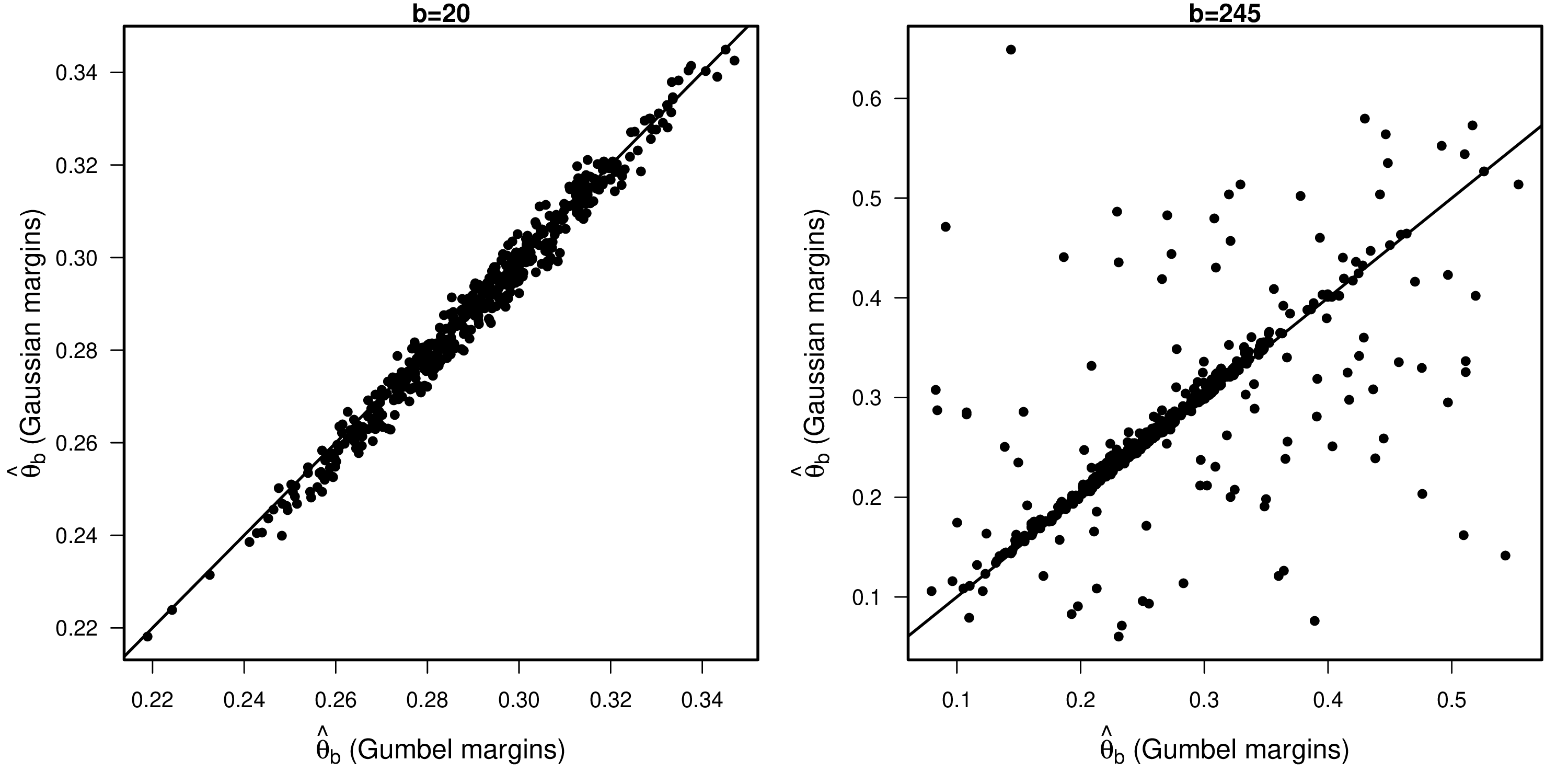}
\caption{\label{fig:Gum_vs_normal}
Estimates $\hat{\theta}_{AT}$ based on data simulated from a moving maxima process with $\alpha=(1/4,1/4,1/4,1/4)$ and $\theta=1/4$: 
Gaussian margins against Gumbel margins.   Left: $b=20$.  Right: $b=245$.}
\end{figure}

We simulate 500 sequences of length $m=4,900$ and estimate $\theta_b$ using the semiparametric estimators (disjoint and sliding blocks) and $\hat{\theta}_{AT}$ and $\hat{\theta}_G$ for $b=20,70,245$ ($n=245,70,20$).
To examine the impact of marginal distribution we apply $\hat{\theta}_{AT}$ and $\hat{\theta}_G$ after transformation to Gumbel, Gaussian and exponential margins.
In fact the marginal distribution has a relatively small effect on the overall performance of $\hat{\theta}_{AT}$ and $\hat{\theta}_{G}$ so we present results only for Gumbel margins.  However, marginal distribution can have an impact on individual estimates.  Figure \ref{fig:Gum_vs_normal} compares the estimates produced by $\hat{\theta}_{AT}$ for a moving maxima process ($X_i=\max_{j=0, \ldots, 3} \{ \alpha_j Z_{i+j} \}$), for $\alpha_j=1/4, j=0, \ldots, 3$, with Gaussian and Gumbel margins.
For small $b=20$ ($n=245$) the agreement is good but for $b=245$ ($n=20$) there is large disagreement for some datasets.

Table \ref{tab:maxAR_05} compares the SP estimators to $\hat{\theta}_{AT}$ and $\hat{\theta}_{G}$.
Naive standard errors are estimated using $n\hat{\theta}_{SP}(n-2)^{-1/2}(n-1)^{-1}$
for the SP estimators and Appendix A for $\hat{\theta}_{AT}$.
For the SP estimators we also estimate adjusted standard errors based on a sandwich estimator (Appendix B) and bootstrap standard errors based on 100 stationary block bootstrap resamples \citep{PR1994} with optimal block length chosen using \cite{PPW2009}, implemented using \cite{boot2014} and \cite{np2008}.
When using disjoint blocks the semiparametric estimator outperforms the parametric estimators approximately to the extent suggested by Figure \ref{fig:rel_eff}.
The use of sliding maxima improves this further.
Comparison of the mean standard errors (SE) with standard deviations (SD)
shows that the (non-bootstrap) standard errors tend to be a little too large, that is, the estimators are less variable than expected.
In general the bootstrap standard errors are more reliable.
As expected, for the estimators based on sliding blocks the naive standard errors are far too small.

\renewcommand{\arraystretch}{1.0}
%
\begin{center}
\begin{table}[!tbp]
 \centering
 \begin{tabular}
{clcccccccc}\hline
\multicolumn{1}{c}{b}&\multicolumn{1}{c}{est}&\multicolumn{1}{c}{$\theta_b$}&\multicolumn{1}{c}{mean}&\multicolumn{1}{c}{RMSE}&\multicolumn{1}{c}{SD}&\multicolumn{1}{c}{SE}&\multicolumn{1}{c}{adj SE}&\multicolumn{1}{c}{boot SE}&\multicolumn{1}{c}{eff}\tabularnewline
\hline
$~20$&d&$0.52$&$0.53$&$0.028$&$0.028$&$0.034$&$0.033$&$0.029$&$$\tabularnewline
$$&s&$$&$0.52$&$0.023$&$0.023$&$0.007$&$0.031$&$0.024$&$1.45$\tabularnewline
$$&AT&$$&$0.53$&$0.037$&$0.037$&$0.050$&$$&$$&$0.56$\tabularnewline
$$&G&$$&$0.52$&$0.038$&$0.038$&$$&$$&$$&$0.54$\tabularnewline
$~70$&d&$0.51$&$0.51$&$0.050$&$0.050$&$0.061$&$0.060$&$0.054$&$$\tabularnewline
$$&s&$$&$0.50$&$0.043$&$0.043$&$0.007$&$0.052$&$0.045$&$1.38$\tabularnewline
$$&AT&$$&$0.51$&$0.071$&$0.071$&$0.091$&$$&$$&$0.50$\tabularnewline
$$&G&$$&$0.50$&$0.075$&$0.075$&$$&$$&$$&$0.45$\tabularnewline
$245$&d&$0.50$&$0.51$&$0.105$&$0.105$&$0.114$&$0.111$&$0.108$&$$\tabularnewline
$$&s&$$&$0.50$&$0.088$&$0.088$&$0.007$&$0.088$&$0.087$&$1.42$\tabularnewline
$$&AT&$$&$0.54$&$0.152$&$0.148$&$0.179$&$$&$$&$0.50$\tabularnewline
$$&G&$$&$0.51$&$0.159$&$0.159$&$$&$$&$$&$0.44$\tabularnewline
\hline
\end{tabular}
 \caption{MaxAR process with $\theta$=0.5. 
Estimators - d: $\hat{\theta}_{d}$;
s: $\hat{\theta}_{s}$;
AT: $\hat{\theta}_{AT}$;
G: $\hat{\theta}_{G}$.
Sampling distribution mean, root mean square error (RMSE) 
and standard deviation (SD), mean standard error (SE), sandwich adjusted 
standard error (adj SE) and boostrap standard error (boot SE) and efficiency (ratio of variances) relative to $\hat{\theta}_d$ (eff).
\label{tab:maxAR_05}} 
\end{table}
\end{center}
\renewcommand{\arraystretch}{1.0}

\vspace*{-1.75cm}

\subsection{SP maxima, blocks, intervals and $K$-gaps estimators of $\theta$}
\label{sec:simb}
\begin{sloppypar}
We compare the performance of the SP maxima estimators to the blocks estimator of \citet{SW1994} and to two of the leading threshold-based estimators: the intervals estimator of \cite{FS2003} and the $K$-gaps estimator of \cite{SD2010}.
The general form of the blocks estimator is $\hat{\theta}_B=\log\hat{G}(u) / b\log\hat{F}(u)$, for some threshold $u$ and block size $b$.
We consider two blocks estimators: the disjoint blocks estimator uses the empirical distribution function of $\{Y_i^d\}$ to estimate $G$, whereas the sliding blocks estimator uses the empirical distribution function of $\{Y_i^s\}$.
\end{sloppypar}

The threshold-based estimators are based on the marginal distribution of the time $T(u_n)=\min\{ k \geq 1 : X_{k+1}>u_n~|~X_1>u_n \}$ between two exceedances of threshold $u_n$.
Under mild conditions, as $n \rightarrow \infty$ the rescaled $K$-gap $\{ 1- F(u_n) \} \max\{T(u_n)-K,0\}$ follows a mixture model: with probability $1-\theta$ the $K$-gap is zero, otherwise it has an exponential distribution with mean $1/\theta$.
The intervals estimator is a moment estimator that (implicitly) uses $K=0$.
The $K$-gaps estimator is a maximum likelihood estimator derived by treating successive $K$-gaps as independent.
\cite{SD2010} note that in practice it is important to use an appropriate value of $K$ and use a model misspecification test to assist this choice, and the choice of threshold.
Based on a simulation study, they find that if $K$ is chosen appropriately, then the $K$-gaps estimator performs better than its competitors: the intervals estimator and the iterative weighted least squares estimator (IWLS) of
\cite{Suveges2007}.

We repeat the simulation study presented in Figure 2 of \cite{SD2010}.
We simulate 1,000 sequences of length $n=30,000$ from each of the processes:
(a) Cauchy AR(1). $X_i = \phi X_{i-1}+Z_i$ with $\phi = 0.7$ and $Z_i$ standard Cauchy: $\theta=0.3$;
(b) Pareto AR(2). $X_i = \phi_1 X_{i-1} + \phi_2 X_{i-2} + Z_i$, with $\phi_1 = 0.95, \phi_2 = −0.89$ and $Z_i$ Pareto with tail index 2: $\theta=0.25$;
(c) A Markov chain with Gumbel margins, a symmetric logistic bivariate distribution
for consecutive variables and dependence parameter $r = 2$ \citep{Smith1992}: $\theta \approx 0.33$.
For the intervals and $K$-gaps estimator we use thresholds corresponding to the 0.95, 0.96, 0.97, 0.98 and 0.99 empirical quantiles.
For the SP maxima and blocks estimators we use block sizes 40, 60, 80, 100, 120, 150 and 200.
The blocks estimators require a block size and a threshold to be set.  
To facilitate comparsion of the SP maxima and blocks estimators we use common block sizes and, for a given block size, we use as the threshold the sample median of the disjoint block maxima for the disjoint blocks estimator and the sample median of the sliding block maxima for the sliding blocks estimator.

Comparison of maxima estimators and threshold estimators is complicated by the different nature of the tuning parameters involved: block size for maxima estimators and threshold for threshold-based estimators.
To provide a tentative basis for comparison we appeal to a result from \cite{Smith1987}, who, for distributions in the domain of attraction of the Gumbel distribution, compared the mean squared error of prediction of extreme quantiles resulting from analyses of block maxima and analyses of threshold exceedances.
Smith found that the optimal sample size (number of exceedances) in the latter is almost double the optimal sample size (number of block maxima) in the former.  
Therefore, in displaying the results of the simulation study we use plotting scales that match a proportion of exceedances $p$ with a block size of $2/p$.

\begin{figure}[H]
\centering
\includegraphics[width=\textwidth, angle=0]{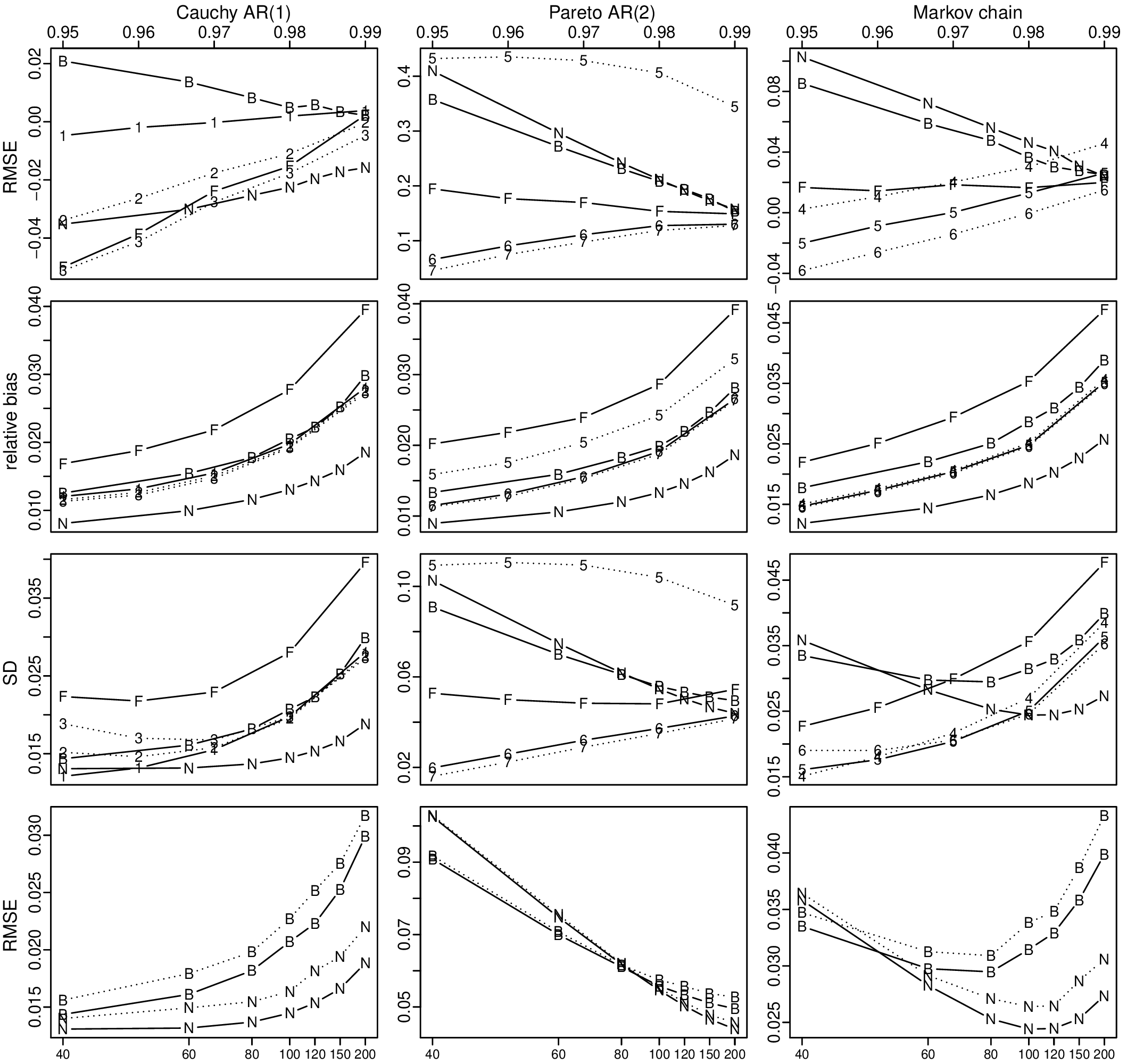}
\caption{\label{fig:SD2010_all}
Top three rows: relative bias, standard deviation and root mean squared error (RMSE) of the sliding blocks SP (labelled N) and blocks (labelled B) estimators, 
the $K$-gaps estimator (labels give the value of $K$; solid line for the optimal $K$, dotted lines otherwise), the intervals estimator (F).  
Bottom row: RMSE of the SP maxima and blocks estimators (dotted lines for disjoint maxima, solid lines for sliding maxima).
Left: Cauchy AR(1); middle: Pareto AR(2); right: symmetric logistic Markov chain.
The upper axis labels give the non-exceedance probability for the threshold-based estimators.  
The lower axis labels give the block size for the SP maxima and blocks estimators.}
\end{figure}

In the top three rows of Figure \ref{fig:SD2010_all} the estimated median relative bias (MRB), standard deviation (SD) and root mean squared error (RMSE) of the sliding blocks versions of the SP maxima and blocks estimators are compared with the threshold-based estimators.
For the $K$-gaps estimator we plot results for the (process-dependent) optimal $K$ determined by \cite{SD2010} (1 for the Cauchy AR(1), 6 for the Pareto AR(2) and 5 for the Markov chain) and for the two values of $K$ closest to the optimal value.

\clearpage

Bias can be attributed to two sources: lack of convergence of $\theta_b$ (or $\theta(u)$) to $\theta$ and bias in estimation of $\theta_b$ (or $\theta(u)$).  The former depends on the process and if convergence is slow then the bias may be a strong determinant of the performance of estimators of $\theta$ even for long sequences of data.
This seems to be the case for the Pareto AR(2), where the RMSE plot mirrors the MRB plot.
For the Cauchy AR(1) (and to a lesser extent for the Markov chain) the biases are smaller so that the relative variabilities of estimators have more influence on the RMSE.
For the Pareto AR(2) the $K$-gaps estimator suffers from relatively large bias and variability if $K$ is chosen to be slightly too small ($K=5$), but for the other two processes the performance of the $K$-gaps estimator is quite insensitive to small deviations from the optimal $K$.
Although choices of block size and threshold make direct comparison difficult, the SP maxima estimators are competitive with the threshold-based estimators.
They have relatively large bias for small block sizes but their low sampling variability results in a relatively small RMSE for larger block sizes.
The SP maxima estimator has lower SD than the blocks estimator, but, particularly for the smaller block sizes, a larger MRB.
The SP maxima estimator uses the data to set local thresholds, whereas the blocks estimator has a constant threshold, which here we have set at the median of the local thresholds.  In this instance, the net effect is that the SP estimator trades a reduction in SD for an increase in MRB.

In the bottom row of Figure \ref{fig:SD2010_all} the disjoint blocks and sliding blocks version of the SP maxima and blocks estimators are compared.
The main advantage of using sliding blocks is a reduction in SD.  
For the Cauchy AR(1) and the Markov chain this effect is apparent in the RMSE.
However, for the Pareto AR(2), where bias dominates, the improvement in RMSE is minimal.

\begin{figure}[h]
\centering
\includegraphics[width=\textwidth, angle=0]{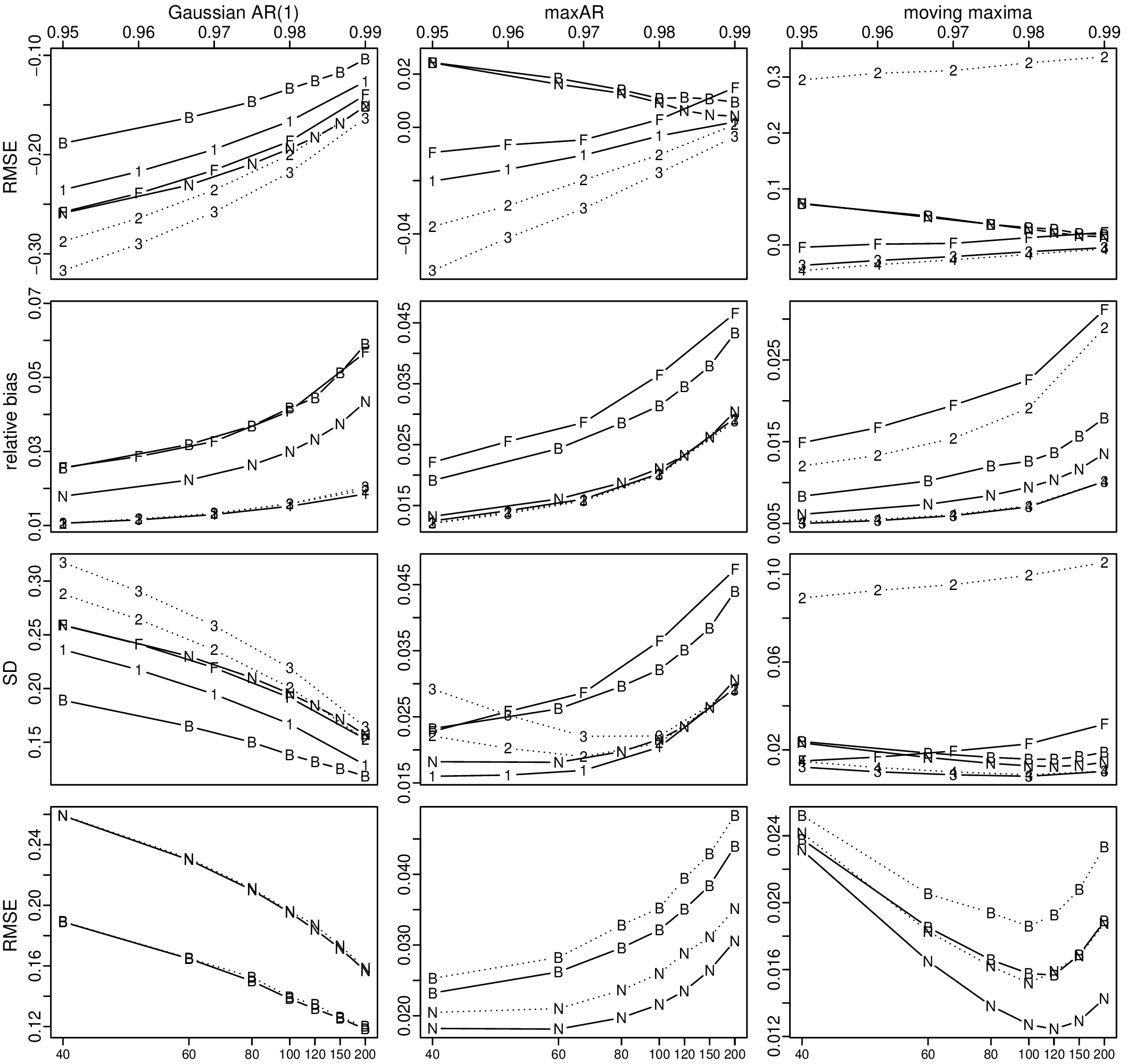}
\caption{\label{fig:sim_extra}
Top three rows: relative bias, standard deviation and root mean squared error (RMSE) of the sliding blocks SP (labelled N) and blocks (labelled B) estimators, 
the $K$-gaps estimator (labels give the value of $K$; solid line for the optimal $K$, dotted lines otherwise), the intervals estimator (F).  
Bottom row: RMSE of the SP maxima and blocks estimators (dotted lines for disjoint maxima, solid lines for sliding maxima).
Left: Gaussian AR(1); middle: maxAR; right: moving maxima.
The upper axis labels give the non-exceedance probability for the threshold-based estimators.  
The lower axis labels give the block size for the SP maxima and blocks estimators.}
\end{figure}

Figure \ref{fig:sim_extra} shows the results of extending the study to three more processes: a Gaussian AR(1) process: (d) $X_i = \alpha X_{i-1}+\epsilon_i$, where  $\{\epsilon_1\}$ are independent N$(0,1-\alpha^2)$, $X_0 \sim N(0,1)$ and we assume $|\alpha|<1$ for second-order stationarity.
This process exhibits serial dependence but limiting extremal independence because 
$\theta=1$ \citep[chapter 4]{LLR1983};
(e) the maxAR process of Section \ref{sec:max_sim}: $\theta=0.5$;
(f) a moving maxima process \citep{Deheuvels1983}:
$X_i = \max_{j=0,\ldots,p} \{ \alpha_j Z_{i+j}\}$, 
where $\alpha_0>0, \alpha_p>0$ and $\alpha_j \geq 0$, for $j=1,\ldots, p-1$, with
$\sum_{j=0}^p \alpha_i = 1$.
$\theta=\max_{i=0,\ldots,p}(\alpha_i)$.  
We consider the case $\alpha=(0.3,0.2,0.2,0.3)$ \citep{ANT2000}
so that $\theta=0.3$.

The findings echo those from Figure \ref{fig:SD2010_all}.
The SP estimators are competitive with the threshold estimators and the $K$-gaps estimator only performs better than the other estimators if $K$ is selected appropriately.
In the maxAR and moving maxima examples the SP estimator fares no worse than the blocks estimator in terms of MRB and better in terms of SD.
In the Gaussian AR(1) case all estimators underestimate the limiting value $\theta=1$ to the extent that bias dominates the RMSE.
The blocks estimators have less bias than the other estimators and therefore have the lowest RMSE of all the estimators.  

\section{Example: Newlyn sea-surges}
\label{sec:newlyn}
\begin{figure}[h]
\centering
\includegraphics[width=0.75\textwidth, angle=0]{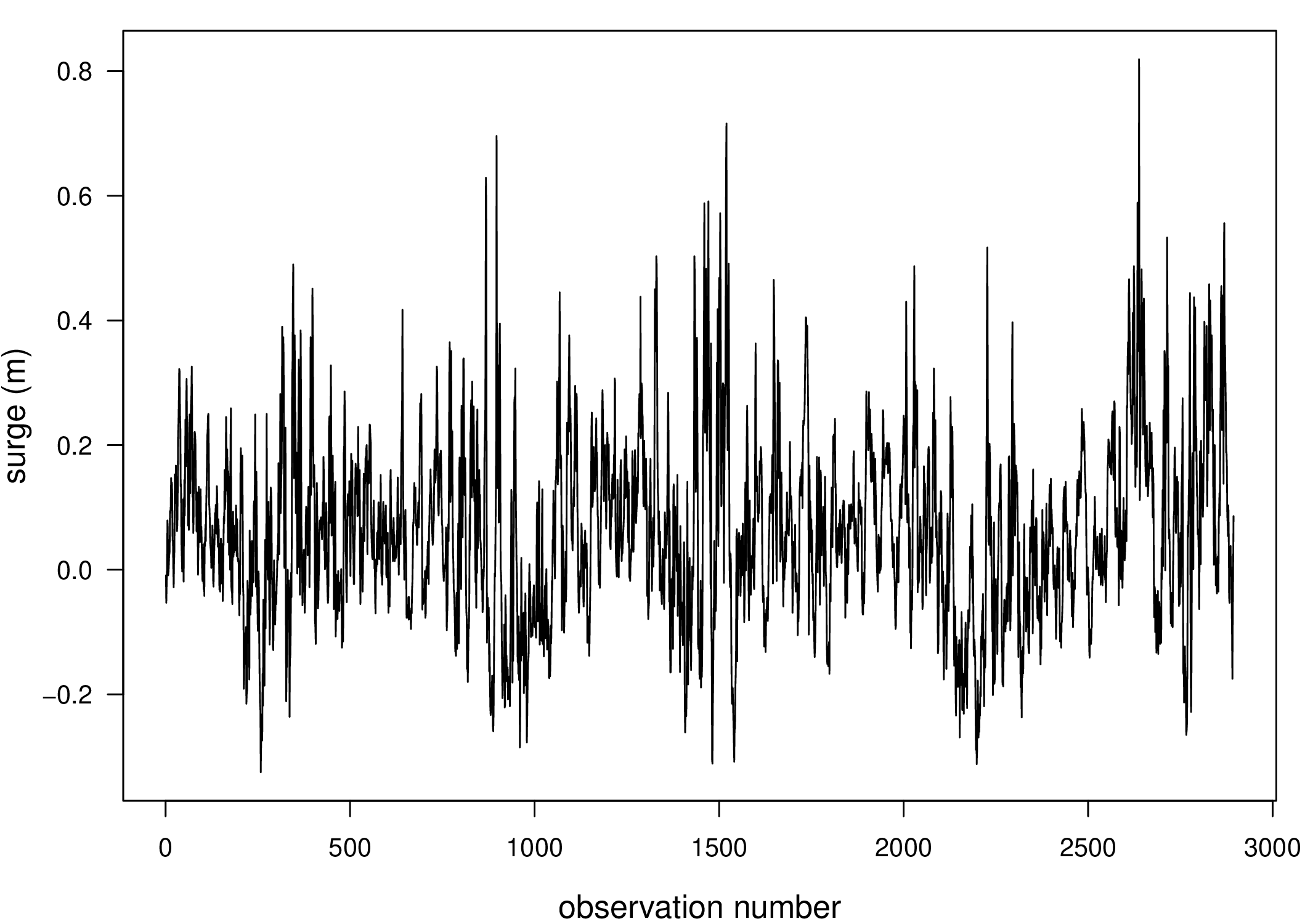}
\caption{\label{fig:newlyn} Time series plot of 2894 maximum sea-surges measured at Newlyn, Cornwall, UK over the period 1971--1976.
The observations are the maximum hourly sea-surge heights over contiguous 15-hour time periods.}
\end{figure}

Figure \ref{fig:newlyn} shows a series of 2894 measurements of sea-surge heights taken just off the coast at Newlyn, Cornwall, UK, over the period 1971--1976.
The data are the maximum hourly surge heights over periods of 15 hours (see \cite{Coles1991}).
\cite{FW2012} used several estimators, including the parametric maxima estimator of \cite{Gomes1993}, to estimate the extremal index of the underlying process using several estimators.
We use $\hat{\theta}_{SP}$ to estimate the extremal index of this series, based on series of disjoint and sliding block maxima.  
We also fit a GEV distribution to the block maxima in order to make inferences about extreme quantiles of the marginal distribution of sea-surge heights at Newlyn.

\clearpage

\begin{figure}[h]
\centering
\includegraphics[width=\textwidth, angle=0]{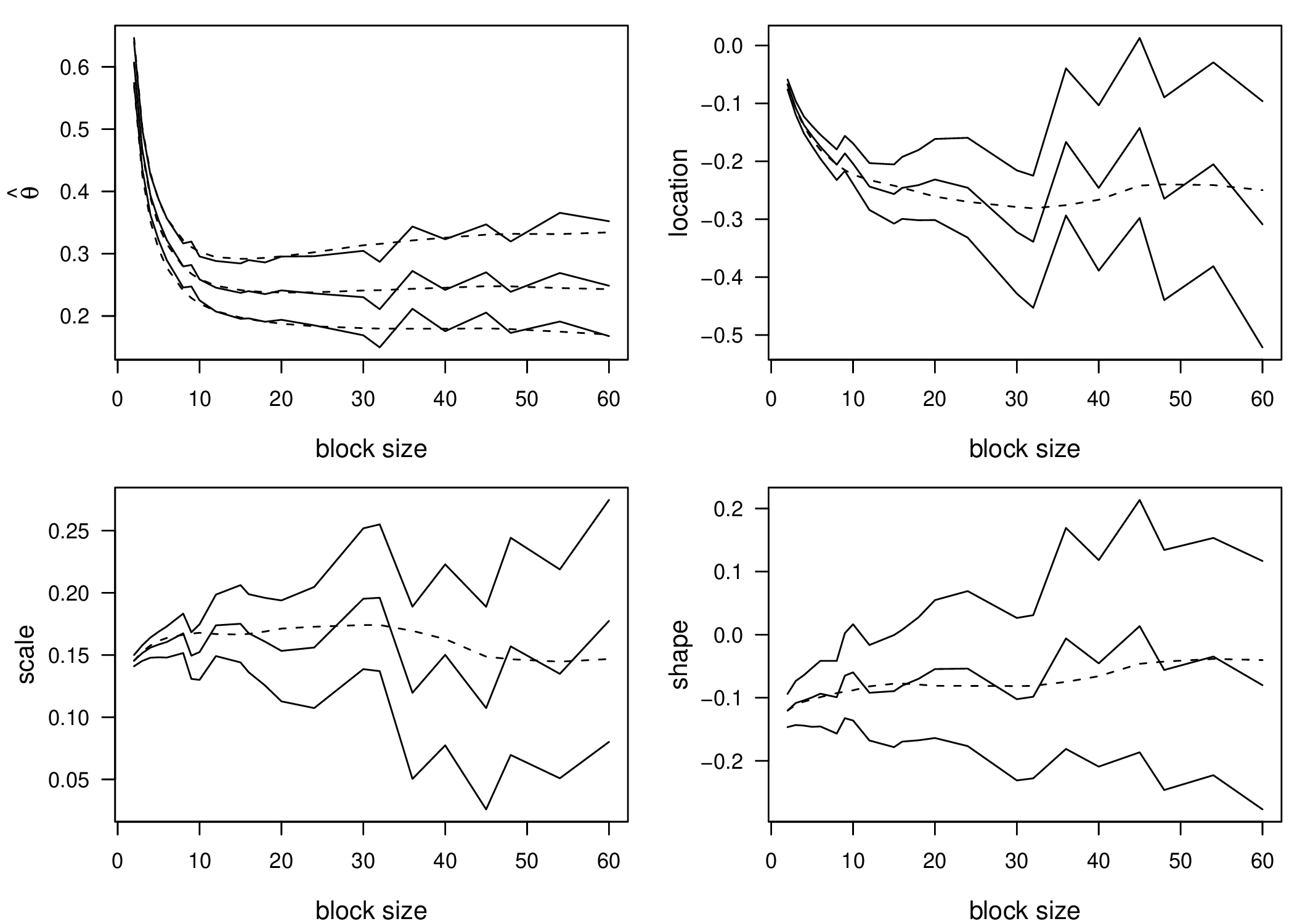}
\caption{\label{fig:newlyn_pars}
Block size selection for the Newlyn data.  Top left: estimates and 95\% confidence intervals for $\theta$ based on disjoint maxima (solid lines) and sliding maxima (dashed lines).  Other plots: estimates of marginal ($b=1$) GEV parameters based on disjoint block maxima (solid lines) and sliding maxima (dashed lines).  Symmetric  95\% confidence intervals are also given for the disjoint maxima.
}
\end{figure}

The top left plot in Figure \ref{fig:newlyn_pars} shows $\hat{\theta}_{SP}$ against block size $b$ based disjoint and sliding maxima.
Also given are 95\% confidence intervals for $\theta$, based on (vertically-scaled) adjusted log-likelihoods, see \citet[page 182]{CB2007}.
The other plots in Figure \ref{fig:newlyn_pars} show maximum likelihood estimates for GEV distributions fitted to the disjoint and sliding maxima, with, for the disjoint maxima only, symmetric 95\% confidence intervals.   As the location and scale of the GEV distribution depend on $b$ we have plotted estimates of the GEV parameters 
implied for a block size of 1, i.e. the marginal distribution of the data.

For these data $b=20$ is reasonable.
Table \ref{tab:newlyn} shows estimates, standard errors and 95\% confidence intervals for $\theta$ using this block size.  
The bootstrap estimates result from the approach detailed in Section \ref{sec:max_sim} using 10,000 resamples.
The accuracy of bootstrap confidence intervals can depend on the parameter scale chosen.
Following \citet[Section 5.2]{DH1997} we seek a monotone {\it variance-stabilizing} transformation $h(\theta)$, with the property that 
${\rm var}[h(\hat{\theta}_{SP})]$ is approximately constant with respect to $h(\theta)$. 
It is also often the case that bootstrap estimates of $h(\theta)$ are closer to being normally distributed than estimates of $\theta$.
From the start of Section \ref{sec:SP} we have ${\rm var}(\hat{\theta}_F) \propto \theta^2$, which suggests that we use $h(\theta)=\log \theta$.
The expression for ${\rm var}(\hat{\theta}_F)$ is not correct in practice, but it may suggest an effective variance-stabilizating transformation.
We construct bootstrap confidence intervals for $\log \theta$ and then transform them back to the $\theta$-scale.
Basic confidence intervals are given in Table \ref{tab:newlyn}, but as the bootstrap distributions of $\log \hat{\theta}_{SP}$ are indeed very close to being normally distributed these intervals are very similar to normal intervals.
Bootstrap bias-adjustment results in a slightly smaller point estimate of $\theta$.

Following \cite{Gomes1993}, \cite{FW2012} used the larger block size $\sqrt{m} \approx 54$, obtaining an estimate of 0.282 with a standard error of 0.206.
For this block size the $\hat{\theta}_{SP}$ compares favourably, with estimates (and adjusted standard errors) of 0.269 (0.044) using disjoint blocks and 0.245 (0.040) using sliding blocks.  The bootstrap standard errors are 0.047 and 0.039 respectively.
\cite{FW2012} also use the intervals estimator of \cite{FS2003}, based on a threshold of 0.3m selected using a mean residual life plot, obtaining 0.223 (0.050).  
The standard errors in Table \ref{tab:newlyn} suggest that, at least for these data, $\hat{\theta}_{SP}$ is competitive with the intervals estimator when both approaches are allowed to select their tuning parameter (block size for $\hat{\theta}_{SP}$ and threshold for the intervals estimator) using the observed data.

\renewcommand{\arraystretch}{1.1}
\begin{table}[h]
\centering
\begin{tabular}{llccc}
& & $\hat{\theta}$ & SE($\hat{\theta}$) & \multicolumn{1}{c}{95\% CI} \\ \hline
\multirow{3}{*}{disjoint} & naive & 0.241 & 0.020 & (0.204, 0.283) \\
& adjusted & 0.241 & 0.026 & (0.194, 0.295) \\
& bootstrap &  0.219 & 0.027 & (0.179, 0.265) \\ \hline
\multirow{2}{*}{sliding} & adjusted & 0.238 & 0.028 & (0.188, 0.296) \\
& bootstrap & 0.213 & 0.023 & (0.180, 0.251) \\ \hline
\end{tabular}
\caption{\label{tab:newlyn} Estimates, standard errors and 95\% confidence intervals for the extremal index $\theta$ of the Newlyn data using (disjoint or sliding) blocks of size of 20.  Naive: independence log-likelihood; adjusted: adjusted log-likelihood; bootstrap: stationary block bootstrap.}
\end{table}
\renewcommand{\arraystretch}{1}

\begin{figure}[h]
\centering
\includegraphics[width=0.75\textwidth, angle=0]{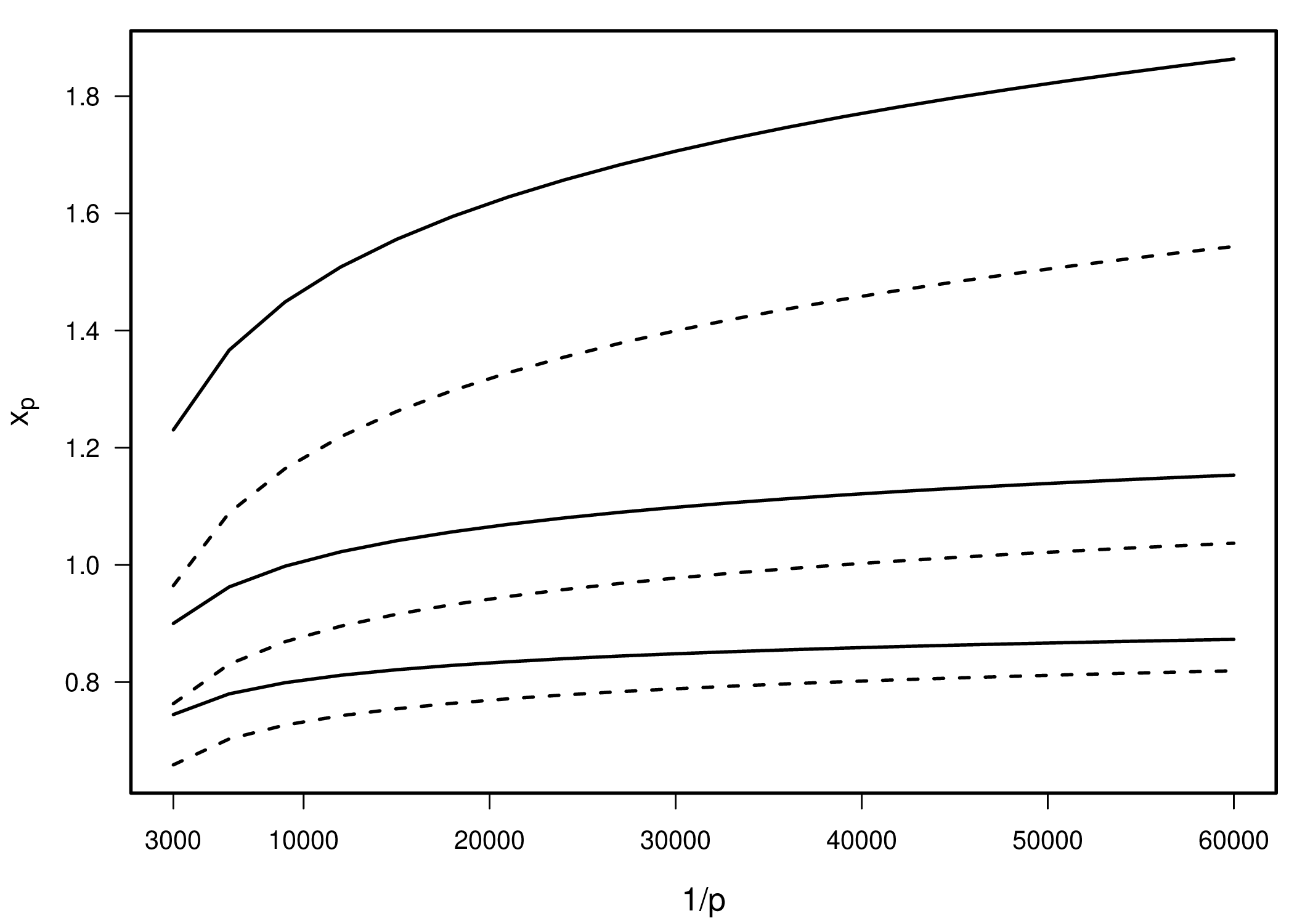}
\caption{\label{fig:newlyn_marg}
Estimates and 95\% confidence intervals for the $100(1-p)$\% marginal quantile $x_p$ against $1/p$.  Solid lines: inferring $\theta$ using $\hat{\theta}_{SP}$ based on sliding maxima.  Dashed lines: using $\theta=1$.}
\end{figure}

Figure \ref{fig:newlyn_marg} shows estimates and 95\% confidence intervals of high quantiles of the marginal distribution of sea-surge height.
The sum of the adjusted log-likelihood for $\theta$ based on sliding maxima and the log-likelihood for the GEV parameters based on disjoint maxima is profiled with respect to the desired quantile.
The estimates (and standard errors) of the GEV parameters $\mu_\theta, \sigma_\theta$ and $\xi$ are 0.192 (0.012), 0.130 (0.0085) and $-0.0546$ (0.056) respectively.
The underestimation that would result from assuming that $\theta=1$ is clear.

\section{Discussion}
\label{sec:discussion}
The semiparametric maxima estimators proposed in this paper improve substantially the existing maxima methods of estimating the extremal index, to the extent that they are competitive with threshold methods.  
The simulation studies in Section \ref{sec:sim} showed that there is benefit to using sliding blocks rather than disjoint blocks.
Apart from the point estimates in Figure \ref{fig:newlyn_pars} we did not use sliding blocks for the GEV analysis in Section \ref{sec:newlyn}.  
As noted by \cite{FP2005}, who employ an approach that is similar to sliding blocks, further research is required to determine how best to provide estimates of uncertainty from analyses based on sliding blocks.

If the main extreme value analysis is threshold-based then, once the threshold has been set, the intervals estimators and the IWLS estimator do not require 
another tuning parameter to be specified.  In contrast the $K$-gaps estimator and the SP maxima estimators do.  However, \cite{SD2010} shows that there is potential benefit in choosing $K$ empirically, jointly with the threshold.
Future work could seek to optimize the choice of block size $b$ empirically.

\section*{Acknowledgments}
I thank Chris Ferro for comments on \cite{Northrop2005}, Lee Fawcett for help with the Newlyn data and Nigel Williams and Marianna Demetriou for project work they carried out in this area.  I am grateful to an associate editor and two anonymous referees: their comments and suggestions have improved the original manuscript.

\section*{Appendix A: asymptotic efficiencies of semiparametric and parametric estimators using disjoint blocks}
The semiparametric estimator $\hat{\theta}_{SP}$ is based on a sample $V_1,\ldots, V_n$ treated as randomly sampled from an exponential distribution with mean $1/\theta$. 
The log-likelihood for a single observation $v$ is 
$l(\theta) = \log \theta +\theta v.$
Thus, the asymptotic precision of $\hat{\theta}_{SP}$ is $-l''(\theta)=1/\theta^2$.

The parametric estimator $\hat{\theta}_{AT}$ of \cite{ANT2000} treats as independent two random samples, each of size $n$.   Sample 1 is from a GEV($\mu,\sigma,\xi$) distribution and sample 2 is from a GEV($\mu_\theta,\sigma_\theta,\xi$) distribution.  
Here we take $n=1$.
Let $I(\mu,\sigma,\xi)$ denote the Fisher information matrix for a sample of size $1$ from a GEV($\mu,\sigma,\xi$) distribution.
This matrix can be inferred from \cite{PW80}, who use a shape parameter $k=-\xi$. 
The calculation of the asymptotic precision of $\hat{\theta}_{AT}$ requires that $\xi>-1/2$ \citep{Smith1985}.

Let $I_1$ and $I_2$ denote the respective Fisher information matrices for the parameter vector $(\theta,\mu,\sigma,\xi)$ from samples 1 and 2. 
$I_1$ is given by
\[ I_1 = \left( {\begin{array}{cc}  0 & \bm{0}  \\  \bm{0}^T & I(\mu,\sigma,\xi)  \\  \end{array} } \right), \]
where $\bm{0}=(0,0,0)$.
Let $\bm{\psi}=(\mu_\theta,\sigma_\theta,\xi)$ and $\bm{\eta}=(\theta,\mu,\sigma,\xi)$ and $\Delta_{ij} = \partial \bm{\psi}_i / \partial \bm{\eta}_j$.  $I_2$ is given by 
$I_2 = \Delta^T I(\mu_\theta,\sigma_\theta,\xi) \Delta.$
The total information is $I=I_1+I_2$, giving
\[ I = \left( {\begin{array}{cc}  1/\theta^2 & \bm{w}  \\  \bm{w}^T & I_{GEV} \\  \end{array} } \right), \]
where $\bm{w}$ is a vector with non-zero entries and $I_{GEV}$ is the total Fisher information for $(\mu,\sigma,\xi)$.
Block inversion of $I$ gives the asymptotic precision of $\hat{\theta}_{AT}$ as
\[ {\rm prec}(\hat{\theta}_{AT}) = 1/\theta^2 - \bm{w} I^{-1}_{GEV} \bm{w}^T > 1/\theta^2 = {\rm prec}(\hat{\theta}_{MLE}), \]
the inequality following because $I^{-1}_{GEV}$ is positive definite.

\section*{Appendix B: estimating the sampling variances}
The sandwich estimator of the sampling variance of $\hat{\theta}$ is 
${\cal J}(\hat{\theta})^{-1} \hat{{\cal V}}(\hat{\theta}) {\cal J}(\hat{\theta})^{-1}$,
where the observed information ${\cal J}(\theta)=n/\theta^2$ and 
$\hat{{\cal V}}(\theta)$ is an estimate of the variance of the score function.
Using the notation defined in Section \ref{sec:SP} the log-likelihood is
\[ l(\theta) = \sum_{i=1}^n \left( \log\theta - \theta \, \hat{V}_i \right). \]
The score function is 
\[ U(\theta) = \sum_{i=1}^n U_i(\theta) = \sum_{i=1}^n \left(\theta^{-1} - \hat{V}_i \right) 
= \theta^{-1} \sum_{i=1}^n \left( 1-\theta \, \hat{V}_i \right). \]
The variance ${\cal V}(\theta)={\rm var}\{U(\theta)\}$ of the score satisfies
\beqn
\theta^2 \, {\cal V}(\theta)&=& \var\left\{ \sum_{i=1}^n \left( 1-\theta \, \hat{V}_i \right) \right\}, \nonumber \\
&=& \sum_{i=1}^n \var \left( 1 - \theta \, \hat{V}_i \right) 
+ 2 \sum_{j=2}^n \sum_{i=1}^{j-1} \cov \left( 1-\theta \, \hat{V}_i, 1-\theta \hat{V}_j \right). \label{eqn:score}
\eeqn
The first term of (\ref{eqn:score}) is estimated by $\sum_{i=1}^n (1-\hat{\theta} \, \hat{V}_i)^2$.  
To estimate the covariance in the second term 
we ignore the possibility that either $\{Y_i<l_i\}$ or $\{Y_j<l_j\}$, as their contributions to the covariance are negligible.
Let
\beqnn
\hat{V}_i \approx -b \log \left\{ \frac{1}{m-b+1} (S_i+T_i) \right\}, \\
\hat{V}_j \approx -b \log \left\{ \frac{1}{m-b+1} (S_j+T_j) \right\},
\eeqnn
where
\beqnn
S_i &=& \sum_{k \notin B_i \cup B_j} I(X_k \leq Y_i), \qquad S_j = \sum_{k \notin B_i \cup B_j} I(X_k \leq Y_j), \\
T_i &=& \sum_{k \in B_j \cap B_i^c} I(X_k \leq Y_i), \qquad T_j = \sum_{k \in B_i \cap B_j^c} I(X_k \leq Y_j).
\eeqnn
In the following we make the simplifying assumption that $\{ X_k, k \in B_i \} \dperp \{ X_k, k \in B_j \}$, for $i \neq j$, that is, data from distinct blocks are independent.
Under this assumption, if disjoint blocks are used then
$S_i \dperp S_j$, $S_i \dperp T_j$ and $S_j \dperp T_i$, because in each case  
a block of $X$s, and/or the maximum of these $X$s, are compared to $X$s from two other disjoint blocks.
Then
\beqnn
\cov\!\left( 1-\theta \hat{V}_i, 1-\theta \hat{V}_j \right) 
&=& \theta^2 \cov (\hat{V}_i, \hat{V}_j), \\
&\approx&  \theta^2 b^2 \cov \left\{  
\frac{1}{m\!-\!b\!+\!1} (S_i\!+\!T_i)\!-\!1, 
   \frac{1}{m\!-\!b\!+\!1} (S_j\!+\!T_j)\!-\!1
\right\}, \\
&=& \frac{\theta^2 b^2}{(m-b+1)^2} \, \cov(T_i, T_j),
\eeqnn
where we have used $\log x \approx x-1$ for $x \approx 1$.
If $\{Y_i>Y_j\}$ then $T_i=b$ and $T_j<b$ and if $\{Y_i<Y_j\}$ then $T_i<b$ and $T_j=b$.
Thus, $(T_i-b)(T_j-b)=0$ and 
\beqnn
\cov(T_i, T_j)&=&\cov(T_i-b,T_j-b), \\
&=& \E\left[ (T_i-b)(T_j-b) \right] - \E(T_i-b) \E(T_j-b), \\
&=& -\left[ \E(T)-b\right]^2, 
\eeqnn
where $\E(T)=b P(X \leq Y) = b^2 \theta/(b\theta+1)$.
As $n(n-1)/2$ pairs of blocks contribute to the second term of (\ref{eqn:score}) it is estimated by $-n (n-1) \hat{\theta}^2 b^4 / (m-b+1)^2 (b\hat{\theta}+1)^2$.

For sliding blocks we note that the second term of (\ref{eqn:score}) contains contributions from pairs of blocks that overlap and $(n-b)(n-b+1)/2$ pairs that do not.  
For the latter the total contribution is estimated by 
$-(n-b)(n-b+1) \hat{\theta}^2 b^4 / (m-b+1)^2 (b\hat{\theta}+1)^2$.
We estimate the total contribution of the former by 
$2 \sum_{k=1}^{b-1} \sum_{i=1}^{n-k} U_i(\hat{\theta}) U_{i+k}(\hat{\theta})$.
Thus, the estimators of ${\cal V}(\theta)$ using disjoint and sliding blocks are respectively
\[ \hat{{\cal V}}_d(\hat{\theta}) = \hat{\theta}^{-2} \left\{ \sum_{i=1}^n (1-\hat{\theta} \, \hat{V}_i)^2 
-\frac{n (n-1) \hat{\theta}^2 b^4}{(m-b+1)^2 (b\hat{\theta}+1)^2} \right\} \]
and
\[ \hat{{\cal V}}_s(\hat{\theta}) = \hat{\theta}^{-2} \left\{  \sum_{i=1}^n (1-\hat{\theta} \, \hat{V}_i)^2 
+ 2 \sum_{k=1}^{b-1} \sum_{i=1}^{n-k} U_i(\hat{\theta}) U_{i+k}(\hat{\theta})
- \frac{(n-b)(n-b+1) \hat{\theta}^2 b^4}{(m-b+1)^2 (b\hat{\theta}+1)^2} \right\}. \]

In practice the contribution to the score function from the largest block maximum $Y_{(n)}$ is non-random, because 
$\hat{V}_{(n)}=-b \log[(m-b)/(m-b+1)]$.  We adjust for this by removing from $\hat{{\cal V}}_d(\hat{\theta})$ and $\hat{{\cal V}}_s(\hat{\theta})$ contributions from $\hat{V}_{(n)}$.

\section*{Appendix C: $\theta_b$ for two processes}
In the following $X_i, i=1,2,\ldots$ are independent unit Fr\'{e}chet random variables with $P(X \leq x)$ = exp$(-1/x)$, for $x>0$.
Thus, $F^b(u_b)=1/2$ implies that $u_b=b/\log2$.

{\it The maxAR process}.
Following \cite[chapter 10]{BGST2004}
\beqnn
G(x)&=&P(X_1 \leq x, \ldots, X_b \leq x), \\
&=&P(X_1\leq x, \theta Z_2 \leq x, \ldots, \theta Z_b \leq x), \\
&=& \exp\left\{ - \left[ 1+\theta(b-1)\right]/x \right\}. 
\eeqnn
Therefore, $\theta_b=-\log G(u_b)/\log 2 = \theta+(1-\theta)/b$.

{\it The moving maxima process}.  
Let $\alpha_i^+=\max(\alpha_0, \ldots, \alpha_i)$ and 
$\alpha_i^-=\max(\alpha_p, \ldots, \alpha_i)$.
Then, for $ b \geq p$,
\beqnn
&&\hspace*{-0.4cm} G(x) = P(X_1 \leq x, \ldots,X_b \leq x), \\
&=& \!\!P\!\left(\!\!Z_1 \!\leq\! \frac{x}{\alpha_0^+}, \ldots, Z_p \!\leq\! \frac{x}{\alpha^+_{p-1}},
Z_{p+1} \!\leq\! \frac{x}{\alpha_p^+}, \ldots, Z_b \!\leq\! \frac{x}{\alpha_p^+}, 
Z_{b+1} \!\leq\! \frac{x}{\alpha_1^-}, \ldots, Z_{p+b} \!\leq\! \frac{x}{\alpha_p^-}\!\!\right)\!\!, \\
&=& \exp\left\{-\left( \sum_{i=0}^{p-1} \alpha_i^+ + (b-p)\alpha_p^+  + \sum_{i=1}^p \alpha_i^- \right) \Bigg/ x \right\}.
\eeqnn
Therefore, 
\beqnn
\theta_b &=& \alpha_p^+ + \frac1b \left( \sum_{i=0}^{p-1} \alpha_i^+   + \sum_{i=1}^p \alpha_i^- - p\alpha_p^+ \right)
= \theta + c/b,
\eeqnn
where $1-\theta \leq  c \leq p\theta$.  The lower bound is achieved if $j=\mbox{argmax}_i\{\alpha_i\} \notin \{0, p\}$ and $\alpha_i=(1-\theta)/p$ for $i \neq j$, or if $\{ \alpha_i\}$ are monotonic in $i$, and the upper bound when $\alpha_0=\alpha_p=\theta$.

\end{document}